\documentclass[aps,prl,twocolumn,superscriptaddress,floatfix]{revtex4-2}
\usepackage[english]{babel}
\usepackage[dvips]{graphicx}
\usepackage{graphics}
\usepackage{amsfonts}
\usepackage{amsmath}
\usepackage{amssymb}
\usepackage{exscale}
\usepackage{eufrak}
\usepackage{afterpage}
\usepackage{upgreek}
\usepackage{color}	%% FOR COLORED NOTES IN DRAFT. COMMENT THIS LINE FOR FINAL VERSION 
\usepackage{braket}	%% For nice BRAKET notation
\usepackage{multirow}	%% For columns spanning multiple rows in Tables
\usepackage{enumitem}
\usepackage{balance}
\usepackage{miller}	%% For nice simplified writing of Miller indices

\usepackage[normalem]{ulem}    %% To use \uline for underlining, 
\newcommand{\CRO}{CuRhO$_2$}
 
\let\oldAA\AA
\renewcommand{\AA}{\text{\normalfont\oldAA}}

\begin{document}
%%%%%%%
\title{Large Seebeck coefficient driven by ``pudding mold'' flat band in hole-doped CuRhO$_2$}
%%%%%%%

\author{Amitayush Jha Thakur}
\affiliation{Universit\'e Paris-Saclay, CNRS,  Institut des Sciences Mol\'eculaires d'Orsay, 
			91405, Orsay, France}
\affiliation{Donostia International Physics Center, 20018 Donostia-San Sebastián, Spain}

\author{Maximilian~Thees}
\affiliation{Universit\'e Paris-Saclay, CNRS,  Institut des Sciences Mol\'eculaires d'Orsay, 
			91405, Orsay, France}	

\author{Franck~Fortuna}
\affiliation{Universit\'e Paris-Saclay, CNRS,  Institut des Sciences Mol\'eculaires d'Orsay, 
			91405, Orsay, France} 
			
\author{Emmanouil~Frantzeskakis}
\affiliation{Universit\'e Paris-Saclay, CNRS,  Institut des Sciences Mol\'eculaires d'Orsay, 
			91405, Orsay, France} 
\author{Daisuke Shiga}
\affiliation{Institute of Multidisciplinary Research for Advanced Materials (IMRAM), 
			Tohoku University, Sendai, 980-8577, Japan}
\affiliation{Photon Factory, Institute of Materials Structure Science, High Energy Accelerator Research Organization (KEK), 1-1 Oho, Tsukuba 305-0801, Japan}

\author{Hiromichi Kuriyama}
\affiliation{Department of Advanced Materials Science, The University of Tokyo, Kashiwa 277-8561, Japan}

\author{Minoru~Nohara}
\affiliation{Department of Quantum Matter, Hiroshima University, Higashi-Hiroshima 739-8530, Japan}

\author{Hidenori Takagi}
\affiliation{Department of Physics, The University of Tokyo, Tokyo 113-0033, Japan}
\affiliation{Max Planck Institute for Solid State Research, 70569 Stuttgart, Germany}

\author{Hiroshi~Kumigashira}
\affiliation{Institute of Multidisciplinary Research for Advanced Materials (IMRAM), 
			Tohoku University, Sendai, 980-8577, Japan}
\affiliation{Photon Factory, Institute of Materials Structure Science, High Energy Accelerator Research Organization (KEK), 1-1 Oho, Tsukuba 305-0801, Japan}

\author{Andr\'es~F.~Santander-Syro}
\email{andres.santander-syro@universite-paris-saclay.fr}
\affiliation{Universit\'e Paris-Saclay, CNRS,  Institut des Sciences Mol\'eculaires d'Orsay, 
			91405, Orsay, France} 
%%%
%%%

%%%%%%%

%%%%%%%%%%%%%
\begin{abstract}
	%%%
	We report the measurement, using angle-resolved photoemission spectroscopy, 
	of the metallic electronic structure of the hole-doped thermoelectric oxide
	CuRh$_{0.9}$Mg$_{0.1}$O$_2$. 
	The material is found to have a ``pudding 
    mold'' type band structure, 
	with a nearly flat band edge located near the Fermi level,
	which is thought to be the origin of the thermoelectric behavior of this material. 
	The experimental data match the density functional theory of the undoped parent compound, 
	simply corrected by a rigid shift of the bands.
	Transport calculations based on the observed band structure yield a Seebeck coefficient of $\sim 200 \,\mu$V/K for the undoped parent material, consistent with experimental measurements. 
	Our results show that CuRhO$_2$ is a textbook example of how pure band-structural effects
	can result in a large thermoelectric figure of merit, demonstrating that flat band edges in oxides are a realistic route for the efficient conversion of thermal energy.
    %%%
\end{abstract}
\maketitle
%%%%%%

%%%%%%%%%%%%%%%%%%%%%%%%
%% INTRODUCTION
%%%%%%%%%%%%%%%%%%%%%%%% 
\section{Introduction}
%%%%%%%%%%%%%%%%%%%%%%%%
Materials that can efficiently convert thermal energy into electricity and vice versa, 
i.e. good thermoelectric materials, are desirable for solid-state cooling applications 
and waste heat conversion to usable energy~\cite{mahan_best_1996}. 
Among many classes of materials under consideration for thermoelectric device applications,
such as bismuth chalcogenides~\cite{witting_thermoelectric_2019} 
or lead telluride~\cite{lalonde_lead_2011}, oxides possess the advantages
of thermal stability (hence operability at high temperatures), 
low toxicity and high oxidation resistance~\cite{malakhov_samsonov_1966}.
The thermoelectric performance of a material is characterized by the 
thermoelectric figure of merit $ZT = S^{2}\sigma T/\kappa$, where $S$ is the Seebeck coefficient, 
$\sigma$ is the electrical conductivity, $T$ is the temperature, and $\kappa$ is the thermal conductivity.
A high Seebeck coefficient, along with good electrical conductivity and low thermal conductivity,
is therefore critical for potential applications. 
However, materials with large carrier concentration, and hence a good electrical conductivity like most metals, 
usually have a low Seebeck coefficient and large thermal conductivity~\cite{blatt_introduction_1976}. 
Doping to introduce carriers increases both the electrical and electronic thermal conductivity,
due to the Wiedemann-Franz law~\cite{jones_theoretical_1985}, 
so this strategy to increase the figure of merit requires efforts to reduce the lattice thermal conductivity.
%%

%%%%%%%%
%% Semiclassical Seebeck Coefficient
%%%%%%%%
Recent theoretical works have shown that materials with a nearly flat band within $k_B T$ of Fermi level ($\mu$)
have both a large Seebeck coefficient and an increased electronic conductivity,
suggesting an alternative route to achieve a large figure or merit.
As demonstrated in~\cite{kuroki_pudding_2007}, under the reasonable assumption that 
the quasiparticle lifetime $\tau(\mathbf{k})$ is momentum-independent, 
and considering only the diagonal part of the Seebeck coefficient tensor as $S$, 
one can approximately write
\begin{equation}
     S = \frac{k_{B}}{e}\frac{\sum_{\mathbf{k}}^{'}v^{2}_{B}-v^{2}_{A}}{\sum_{\mathbf{k}}^{'}v^{2}_{B}+v^{2}_{A}},
\label{Eq:Seebeck_approx}
\end{equation}
where $k_B$ is the Boltzmann constant, 
$e < 0$ is the electron charge, 
$\varepsilon(\mathbf{k})$ is the momentum-dependent energy of the conduction band,
$\sum^{'}_{\mathbf{k}}$ is a sum over $\mathbf{k}$ states 
with $|\varepsilon(\mathbf{k})-\mu|<k_{B}T$, and $v_{A}$, $v_{B}$ are respectively 
the group velocities of states just above and below the chemical potential $\mu$. 
The presence of a flat band just above (below) $\mu$ 
which changes to a highly dispersive band below (above) $\mu$, 
i.e $v_{A}^{2}\ll v_{B}^{2}$ ($v_{A}^{2}\gg v_{B}^{2}$), 
results in a large value of $|S| \sim O(k_{B}/|e|) \sim O(100) \mu$V/K 
along with relatively low resistivity coming from the conducting band, which results in a strong enhancement in the thermoelectric figure of merit $ZT$.
This type of band structure is referred to as the ``pudding mold'' type band, 
and has been theoretically investigated in oxides such as Na$_{x}$CoO$_2$~\cite{kuroki_pudding_2007}, 
LaRhO$_3$ and CuRhO$_2$~\cite{Usui_2009} and other materials like FeAs$_2$ 
and PtSe$_2$~\cite{usui_pudding-mold_2014}.
%%%

\begin{figure}[!t]
     \centering
     	\includegraphics[clip, width=\columnwidth]{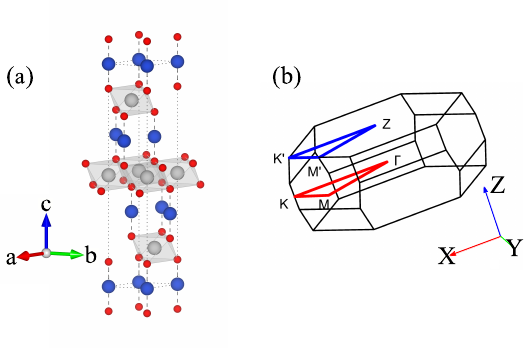}
  	 \caption{\footnotesize{
      		 (a)~The conventional hexagonal unit cell of CuRhO$_{2}$, 
      		 with atoms colored red (O), blue (Cu), and gray (Rh). 
      		 (b)~Brillouin zone (BZ) of \CRO, showing high-symmetry points ($\Gamma, K, M$) and the path used for band structure calculations. 
  		}
  	} 
\label{Fig:CRO-structure}
\end{figure}

%%%
Transition-metal oxides, some of which are both conductive and thermoelectric materials, 
are in principle good candidates for realizing such ``pudding mold'' band scenario, 
thanks to the narrow $d$-orbitals constituting their conduction band 
and the natural occurrence of layered structure which promotes in-plane electrical conductivity 
and minimizes the out-of-plane thermal conductivity~\cite{romanenko_review_2022}.  
In particular, the layered oxide CuRhO$_{2}$ has raised interest 
as thermoelectric material~\cite{4133245,Usui_2009,kurita_correlation_2019}.
It belongs to the $166-R\overline{3}m$ space group with the typical delafossite structure, 
with RhO$_2$ layers stacked between the the triangular Cu layers 
(see Fig.~\ref{Fig:CRO-structure}), 
a semiconducting band gap of $1.9$ eV~\cite{gu_p-type_2014},
and a high Seebeck coefficient of $\sim 200\, \mu$V/K at 300 K~\cite{kurita_correlation_2019}.
Mg substitution of Rh sites leads to the introduction of hole carriers in the material, 
resulting in a transition to a metallic state. The Seebeck coefficient decreases systematically 
with doping in single-crystalline samples, 
but the 10\% Mg-substituted compound CuRh$_{0.9}$Mg$_{0.1}$O$_2$ (CRMO)
still exhibits a relatively large Seebeck coefficient of $\sim 100\, \mu$V/K at 300 K~\cite{kurita_correlation_2019}.  
%%%

%% Figure 1
%%%%

%%
On the other hand, as narrow $d$-bands might also result in strong electron correlations,
another mechanism for the enhancement of the Seebeck coefficient has been proposed 
in transition-metal oxides, namely correlation-induced spin degeneracy, 
as demonstrated e.g. in the case of NaCo$_2$O$_4$~\cite{koshibae_thermopower_2000}.
Therefore, accessing the experimental electronic structure of thermoelectric transition-metal oxides
is essential to disentangle the effects of strong correlations from those of pure band-structural origin, 
and to verify the mechanism driving the enhanced Seebeck coefficient.
%%%
More generally, an experimental demonstration of a ``pudding mold'' band 
in a good thermoelectric material, together with a proof that such band structure is at the
origin of the material's large Seebeck coefficient, has been so-far missing.
%%%

%%%%%%%%%% HERE WE SHOW 
In this paper we report the direct observation, 
through angle-resolved photoemission spectroscopy (ARPES),
of a ``pudding mold'' electronic structure 
of metallic 10\% Mg doped CuRhO$_2$ (CRMO).
We compare our results with density functional theory (DFT) calculations, which confirm the presence of a nearly flat band edge near the Fermi level and yield a theoretical estimate of the Seebeck coefficient that agrees well with the experimental measurements.

%%%%%

%%%%%%%%%% METHODS 
% \section{Materials and methods}
%%%%%%%%%%%%%%%%%%%%%%%%%
 %$CuRhO_2$ crystals 

%Growth% 
Single crystals of CuRh$_{0.9}$Mg$_{0.1}$O$_2$ were synthesized using the self-flux method. First, polycrystalline samples were prepared by reacting a stoichiometric mixture of CuO, Rh$_2$O$_3$, and MgO at 1050 °C for 24 hours with intermediate grinding. The resulting powder was then mixed with CuO at a ratio of CuO:CuRh$_{0.9}$Mg$_{0.1}$O$_2$ = 10:1 in a platinum crucible. The mixture was heated to 1200 °C for 5 hours, then slowly cooled to 1050 °C at a rate of 0.5 °C per hour, followed by rapid cooling to room temperature at a rate of 300 °C per hour. Finally, the residual flux was removed using an aqueous solution of 1 M HNO$_3$. 

%Measurements%
ARPES measurements were performed using synchrotron radiation at the BL2-A (Mushashi) beamline 
in Photon Factory, KEK. The electron analyzer in the setup was Scienta-Omicron SES-2002. Soft X-rays (SX) photons with linear horizontal (LH) polarization—that is, p-polarized with respect to the measurement geometry—in the energy range of 500–620 eV, and vacuum ultraviolet (V-UV) photons at 182 eV with LH polarization were used. 
The CRMO samples were cleaved along $[001]$ 
and measured in UHV conditions with pressure maintained around $5\times10^{-11}$mbar 
at a temperature of $20$~K. 
The Fermi level was set by fitting a Fermi function convoluted with a linear function 
to the momentum integrated energy slices of the energy-momentum measurements. 
%%%

DFT calculations of CuRhO$_2$ electronic structure were performed with the 
QUANTUM ESPRESSO~\cite{giannozzi_quantum_2009} code with a plane wave basis, 
using a polarized general gradient approximation (GGA) exchange-correlation functional 
of Perdew-Burke-Ernzerhof (PBE) from the Standard solid-state pseudopotentials (SSSP) 
efficiency library~\cite{prandini_precision_2018-1,lejaeghere_reproducibility_2016}.
The plane wave kinetic energy cutoff is set to 55 Ry and the charge density kinetic energy cutoff is 440 Ry. 

To simulate 10\% Mg-doped CuRhO$_2$, we performed electronic structure calculations on undoped CuRhO$_2$ with a small modification to its crystal structure. The lattice parameters were set to $a = 3.068$~\AA~and $c = 17.09$~\AA~(with respect to the conventional hexagonal lattice). To simulate the doped material, the lattice parameter $a$ was set to be approximately 0.2\% smaller than the value reported for undoped CuRhO$_2$ in Ref.~\cite{bertaut_sur_1961}. This adjustment accounts for the reduction in the average ionic radius at the Rh site due to Mg doping, as described in Ref.~\cite{kurita_correlation_2019}. The parameter $c$ remains the same as that for the undoped material. The unit cell was then relaxed by allowing the atomic positions to adjust while keeping the cell dimensions fixed. Finally, the calculated bands were aligned with the band measured in the Mg-doped compound by applying a rigid band shift of -0.06~eV from the top of the valence band maximum. This shift provides a first approximation of the electronic structure changes induced by hole doping via Mg substitution. The value of 0.06~eV was determined by observing the shift in the Fermi level in the DOS obtained from a supercell (SC) calculation consisting of 10 Rh sites, in which one Rh atom was replaced by a Mg atom that corresponds to 10\% Mg doping (Supplementary Material, Section S1).
 
%%%
The BoltzTrap package~\cite{madsen_boltztrap._2006} utilizing semi-classical transport equations 
was used to calculate the Seebeck coefficient based on the DFT band structure of the doped material. The calculation of Seebeck coefficient of the undoped material was also done for comparison, based on the DFT calculations of pristine CuRhO$_2$ structure ($a = 3.075$~\AA~and $c = 17.09$~\AA~) without any band shifts.
%%%%%

%%%%%%%%%%%%%%%%%
%%% RESULTS
%%%%%%%%%%%%%%%%%
% \section{Results}
%%%%%
%% Figure 2
\begin{figure}[!t]   
  \centering   
     \includegraphics[clip, width=\columnwidth]{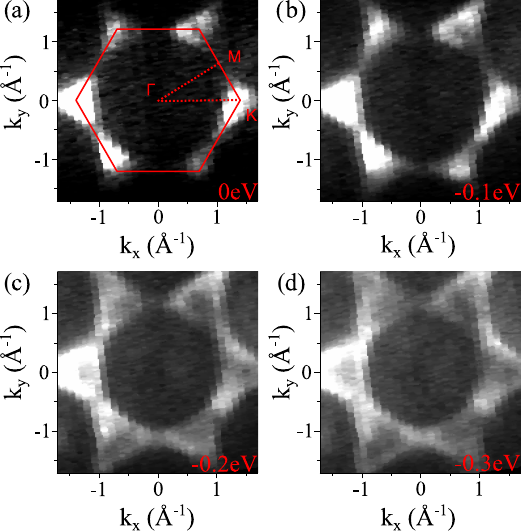}
  \caption{\footnotesize{ 
          Constant energy maps of \CRO~measured at $h\nu = 182$ eV (Maps at SX energy of $h\nu = 542$ eV are presented in Supplementary Material, Section S4). Maps are shown at energies of (a)~$0$ eV, (b)~$-0.1$ eV, (c)~$-0.3$ eV, and (d)$-0.4$ eV relative to the Fermi level. The hexagonal BZ is overlaid in red in (a). 
  	    }
        } 
\label{Fig:CEMs} 
\end{figure}
%%%

%%%
As shown in Fig.~\ref{Fig:CEMs}(a), the Fermi surface of CRMO at the bulk $\Gamma$ plane
is composed of six symmetrical triangular hole-like pockets centered at the high-symmetry $K$-points. Figs.~\ref{Fig:CEMs}(b-d) show the evolution of the constant energy surface at different energies.
%%%

%% Figure 3
\begin{figure}[!t] 
  \centering        
     \includegraphics[clip, width=\columnwidth]{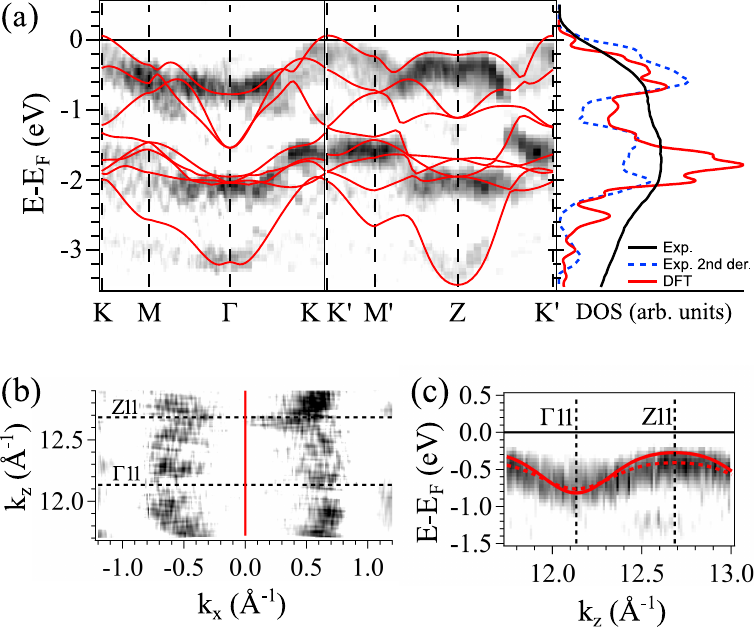}
  \caption{\footnotesize{
  		  (a) Left: ARPES in-plane band dispersion  along $K–M–\Gamma–K$ at 542 eV and $K'–M'–Z–K'$ at 590 eV. Right: Angle-integrated energy dispersion for raw ARPES data (black curves), second derivative intensity (dashed blue curves) and calculated DOS (red curves).
  		  (b) Out-of-plane Fermi surface along $k_z-k_x$ plane,
  		  spanning the high symmetry points $\Gamma11$ ($h\nu = 538$ eV) and $Z11$ ($h\nu = 590$ eV). 
  		  % The photon energy range is $500-620$eV \textcolor{red}{[What polarization?]}.
  		  (c) Energy-momentum dispersion along $k_z$ (normal emission) with $k_x$ along $M\text{--}\Gamma\text{--}M$ direction. Red solid curves represent theoretical bands, while dotted red curves show experimental fits. 
  		  %%%
  		  In panels (b) and (c), the data was obtained using photons 
  		  in the range of $500-620$ eV energy. $\Gamma$11 is calculated to lie at $538$ eV by fitting with the $k_{z}$ measurement and using an inner potential of $23.5$ eV in the three step model approximation~\cite{hufner_photoelectron_1995}.
  		  %%%
  		  }
  		}
\label{Fig:Dispersions} 
\end{figure}

Fig. \ref{Fig:Dispersions}(a) shows the second derivative intensity plots of the ARPES dispersion, calculated along the energy axis. The ARPES data, measured along the high-symmetry paths $K–M–\Gamma–K$ at 542 eV and $K'–M'–Z–K'$ at 590 eV, are overlaid with the corresponding DFT calculations. For reference, the unprocessed (raw) ARPES dispersions can be found in Section S3 of the Supplementary Material. Fig.~\ref{Fig:Dispersions}(b) shows the out-of-plane Fermi surface map,
spanning the $\Gamma$11 (referring to $\Gamma_{<0\,0\,11>}$) and $Z11$ (referring to $Z_{<0\,0\,11>}$) high symmetry points in reciprocal space. The corresponding ARPES energy-momentum dispersion along $k_{z}$ (momenta along out-of-plane z direction),
shown in Fig.~\ref{Fig:Dispersions}(c),
reveals a cosine-like band dispersion with an electron-like minimum around $k_{z}=\Gamma$ 
and a hole-like maximum at $k_{z}=Z$. 
The experimental bandwidth along $k_{z}$ is smaller than the calculated one by a factor of about $0.65$. This can be due to unaccounted correlations of the Rh 4d t$_{2g}$ orbitals in the calculations. This discrepancy in band dispersion along $k_{z}$ is also seen in the in-plane dispersion at $k_{z}=Z $ (Fig. \ref{Fig:Dispersions}(a)), where the valence band maxima around $Z$ is shifted $\sim$ -0.2 eV compared to the calculations. 

We also compare the calculated DOS with the angle-integrated ARPES intensity in Fig. \ref{Fig:Dispersions}(a). The raw angle-integrated intensity exhibits broad peaks centered around –0.7 eV and –1.7 eV that align with the main features of the calculated DOS, but it does not capture the finer peak structures or the precise bandwidths. In contrast, the angle-integrated intensity of the second derivative resolves the finer double-peak feature near the Fermi level, as well as the bandwidths of the bands around –0.7 eV and –1.7 eV. From the ARPES measurements of the dispersion and the angle-integrated intensities, the conduction band bandwidth is determined to be $\sim$ 0.8 eV, which matches the calculations. Overall, the DFT calculations accurately reproduce the experimentally observed electronic structure.

%%%%
 %% Figure 4
\begin{figure}[!t]  
  \centering 
     \includegraphics[clip,width=\columnwidth]{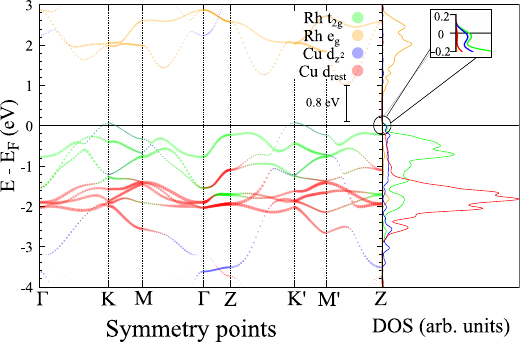}
  \caption{\footnotesize{Left: Orbital-projected bands of CRMO (with the Fermi level set   shifted to -0.06 eV from the top of the valence band) with orbital weights indicated by the size of coloured circles.
  		  Right: Orbital-projected DOS. Inset: A closer view of the DOS near the Fermi level shows the density of Rh t$_{2g}$ orbitals and Cu d$_{z^2}$ orbitals are similar at the Fermi level. Cu d$_{\text{rest}}$ refers to the combined weight of the Cu d orbitals excluding the Cu d$_{z^2}$ orbital. Vertical black line indicates gap between conduction band minima and valence band maxima.
  		}
  	} 
\label{Fig:Orb_res}
\end{figure}
%%%

%%%
As shown in Fig. \ref{Fig:Orb_res}, the calculated gap between the valence band maxima and the conduction band minima is $\sim 0.8$ eV, which is smaller than the experimental value of $\sim$ 1.9 eV for the undoped material. This discrepancy can be attributed to the underestimation of the insulating band gap 
by DFT~\cite{https://doi.org/10.1002/qua.560280846}. Although DFT+U calculations match the experimental gap more closely, the overall dispersion from standard DFT is in closer agreement with experimental band structure data (see Supplementary Material, Section S2).
%%%
The calculations indicate that the conduction band edge of CRMO exhibits distinct orbital characters depending on momentum. At the $K$ and $K'$ points, the conduction band comprises Cu 3$d_{z^2}$ states hybridized with Rh 4d $t_{2g}$ orbitals, whereas at the $\Gamma$ and $Z$ points it is composed entirely of Rh $t_{2g}$ orbitals.
%%%

From the calculated bands, the in-plane group velocities near $K$ (i.e. top of the conduction band) 
are $v_{A}= 0.57$~eV$\AA$ and $v_{B} = 1.00$~eV$\AA$,
while the group velocities near $K'$ are $v_{A} = 0.49$~eV$\AA$ and $v_{B} = 1.36$~eV$\AA$
averaged within an energy window of $k_{B}T \approx 1000$~K around the Fermi level. Assuming that the in-plane Seebeck coefficient in CRMO comes essentially
from electronic states around $K'$,
and using the above values of group velocities above and below $E_F$ around $K'$,
one obtains from Eq.~\ref{Eq:Seebeck_approx},
an estimated value of $S \sim 66 \,\mu$V/K which is of the same order as the experimental value of  
$\sim 100\, \mu$V/K in CRMO measured at 300 K~\cite{kurita_correlation_2019}.
%%

%%Figure 5
\begin{figure}[!t]  
  \centering 
     \includegraphics[clip,width=\columnwidth]{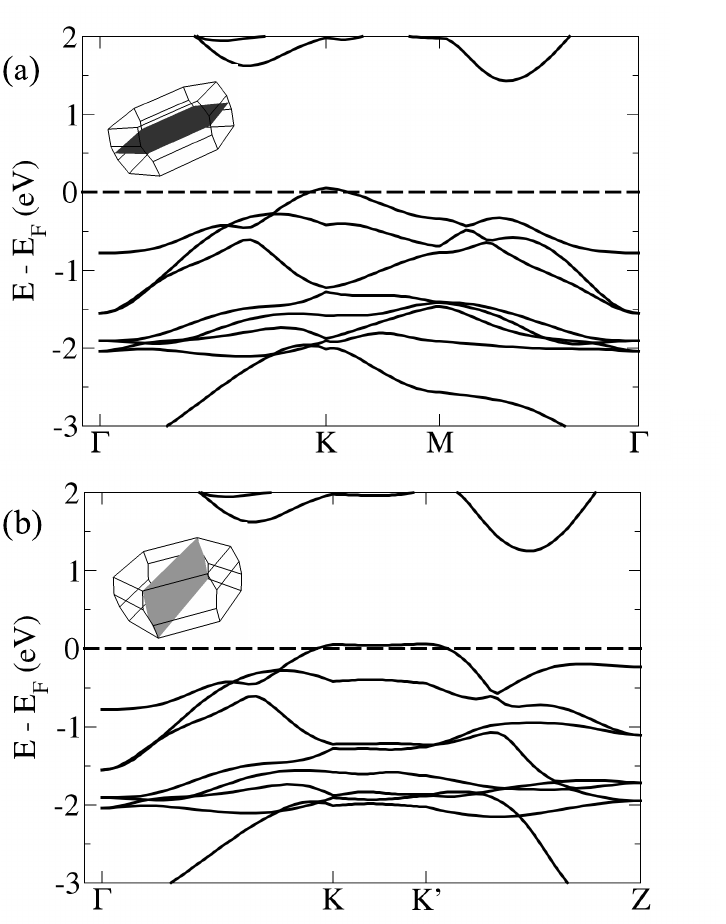}
  \caption{\footnotesize{Band structure plots along (a) $\Gamma\text{--}K\text{--}M\text{--}\Gamma$ and (b) $\Gamma\text{--}K\text{--}K'\text{--}Z$ high-symmetry points. The plane in which these high-symmetry points lie, located within the Brillouin zone, is illustrated in the inset. }
        } 
\label{Fig:pudding_band}
\end{figure}

To better illustrate the ``pudding mold'' nature of the bands, the in-plane bands along $\Gamma–K–M–\Gamma$ and the out-of-plane bands along $\Gamma–K–K'–Z$ for CRMO are plotted in Fig.~\ref{Fig:pudding_band}. For the in-plane bands, we observe that the conduction band maximum lies just above the Fermi level at the $K$ point. This feature results in an enhancement of the Seebeck coefficient as $v_A < v_B$ in the small region around $K$. However, the out-of-plane conduction band has a large flat region between the $K$ and $K'$ points and is a model example of the ``pudding mold'' band structure. This ``pudding mold'' feature leads to a greater enhancement of the Seebeck coefficient in the out-of-plane direction.

%% Figure 6
\begin{figure}[!b]
  \centering
     \includegraphics[width=\columnwidth]{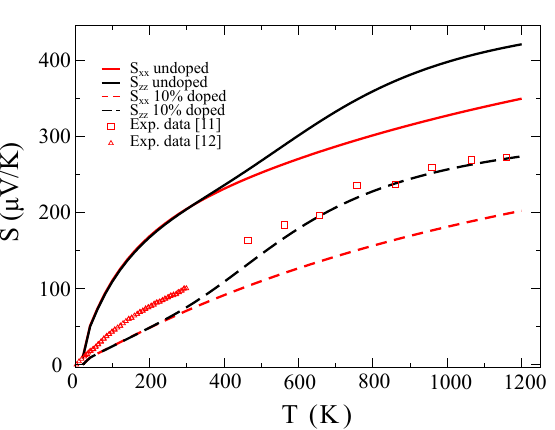}
  \caption{\footnotesize{The $S_{xx}$ and $S_{zz}$ components of the calculated Seebeck coefficient versus temperature level for the undoped sample and the 10\% doped sample. Experimental data from \cite{4133245} and \cite{kurita_correlation_2019} are compared to the calculations of the doped Seebeck coefficient.
  		}
  	} 
\label{Fig:SvT}
\end{figure}
%%%

%%%
Fig.~\ref{Fig:SvT} shows the calculated Seebeck coefficients for the undoped and the 10\% doped material. Undoped \CRO~has a Seebeck coefficient of $\sim 200\, \mu$V/K at 300 K, 
and the 10\% doped sample CRMO
has a Seebeck coefficient
of $\sim 70\, \mu$V/K at 300 K. These values closely match the values measured for polycrystalline samples \cite{shibasaki_transport_2006}. However, they are slightly smaller than the values measured for single crystals \cite{kurita_correlation_2019}. This discrepancy might be due to the effects of electron-phonon interactions observed in \CRO ~single crystals \cite{kurita_correlation_2019}.
Furthermore, the comparisons between the in-plane ($S_{xx}$) and out-of-plane ($S_{zz}$) 
Seebeck coefficient show that, above 300 K, the out-of-plane component of the Seebeck coefficient increases faster than the in-plane component. 
This is ascribed to the flatter dispersion in the $K–K'$ direction.
The high value of $S_{zz}$ $\sim 256\, \mu$V/K at 1000 K for the doped sample 
is very close to the experimental value of $\sim 260\, \mu$V/K reported in Ref.~\cite{4133245}. 
%%%

In conclusion, our ARPES observations of the electronic structure near the Fermi level for CRMO fit well with the calculated band structure and its description in terms of the ``pudding mold'' band model.
Our work provides an experimental basis for the realization of materials with large Seebeck coefficients and thermoelectric figure of merit, originating from a flat conduction band with the band edge lying close to the Fermi level.
%%%%
 
%%%%%%%%%%
\section{Acknowledgements}
%%%%%%%%%%
Work at ISMO was supported by public grants from the 
Agence Nationale de la Recherche (ANR), project Fermi-NESt No. ANR-16-CE92-0018,
by the ``Laboratoire d'Excellence Physique Atomes Lumi\`ere Mati\`ere''
(LabEx PALM ) projects ELECTROX, 2DEG2USE and 2DTROX, 
overseen by the ANR as part of the ``Investissements d'Avenir'' program ANR-10-LABX-0039,
and by the CNRS International Research Project EXCELSIOR.
Work at Tohoku University was supported by 
Grants-in-Aid for Scientific Research (Nos. 16H02115 and 16KK0107) 
from the Japan Society for the Promotion of Science.
Experiments at KEK-PF were performed under the approval of the Program Advisory Committee 
(Proposals 2016G621 and 2018S2-004) at the Institute of Materials Structure Science at KEK.
%%%
%%%
This work was performed using HPC resources from the M\'esocentre computing center 
of CentraleSup\'elec and \'Ecole Normale Sup\'erieure Paris-Saclay supported by CNRS 
and R\'egion \^Ile-de-France (http://mesocentre.centralesupelec.fr/).

%%%%%%%%%%%%
% \printbibliography
%%%%%%%%%%%%
%% \bibliography{CRMO_V13}

%apsrev4-2.bst 2019-01-14 (MD) hand-edited version of apsrev4-1.bst
%Control: key (0)
%Control: author (8) initials jnrlst
%Control: editor formatted (1) identically to author
%Control: production of article title (0) allowed
%Control: page (0) single
%Control: year (1) truncated
%Control: production of eprint (0) enabled
%

%%%%%%%%%%%%%

%%%%%%%%%%%%%

\end{document}